\documentclass[12pt]{article}

\usepackage[latin1]{inputenc}
\usepackage{tikz}
\usepackage{enumitem}
\usetikzlibrary{shapes,arrows,positioning}
\usepackage{verbatim}
\usepackage{fullpage}
\usepackage{amsmath,amsthm,amssymb}

\usepackage{bibspacing}
\usepackage{flexiblemathdisplay}

\theoremstyle{definition}
\newtheorem{lemma}{Lemma}

\newtheorem{claim}[lemma]{Claim}

\numberwithin{lemma}{section}

\newcommand{\poly}{\mathrm{poly}}

\newcommand{\zo}{\{0, 1\}}

\newcommand{\e}{\epsilon}
\newcommand{\eps}{\epsilon}

\renewcommand{\lg}{\log}

\tikzstyle{box} = [rectangle, draw, text width=9em, text
badly centered, node distance=5cm, inner sep=3pt]

\tikzstyle{box4} = [rectangle, draw, text width=4em, text
badly centered, node distance=5cm, inner sep=3pt]

\tikzstyle{box7} = [rectangle, draw, text width=7em, text
badly centered, node distance=5cm, inner sep=3pt]

\tikzstyle{box11} = [rectangle, draw, text width=11em,
text badly centered, node distance=5cm, inner sep=3pt]

\tikzset{arr/.style={->, line width = 5} }

\newcommand{\mycon}{\refstepcounter{lemma}
(\arabic{section}.\arabic{lemma})}

\begin{document}

\begin{titlepage}

\title{Challenges in computational lower bounds\\%(Preliminary version)
}
\author{Emanuele Viola\thanks{Supported by NSF grants CCF-0845003, CCF-1319206.  Email: \texttt{viola@ccs.neu.edu}}}

\maketitle

We draw two incomplete, biased maps of challenges in
computational complexity lower bounds.  Our aim is to put
these challenges in perspective, and to present some
connections which do not seem widely known.

We do not survey existing lower bounds, go through the
history, or repeat standard definitions.  All of this can
be found e.g.~in the recent book \cite{Jukna2012}, or in
the books and surveys
\cite{ShpilkaY10,Lokam09,viola-FTTCS09,She08-survey,SanjeevBarak2007,KuN97,Bei93,Razborov91,BoppanaSipser90,Has87}.

Each node in the maps represents the challenge of proving
that there exists an explicit boolean function that
cannot be computed with the resources labeling that node.
We take explicit to mean NP, thus excluding most or all
of the lower bounds that rely on diagonalization.  An
arrow from node A to B means that resources A can
simulate resources B, and so solving A implies solving B.

%\section{Issues}

%Thus, from a logical point of view, sinks should be
%attacked first.

%But for context we mention that, for most of the
%challenges listed next, if the parameters are made a bit
%less demanding than a lower bound is known.

%Some of the arrows do not seem to have appeared
%explicitly in the literature and so we include their
%proofs.

%keywords
%turing machine, independent source, deterministic randomness extractor, sampling lower bound, complexity of distributions

\thispagestyle{empty}
\end{titlepage}

\section{Circuits with various  gates, correlation, and communication}

%\begin{figure}
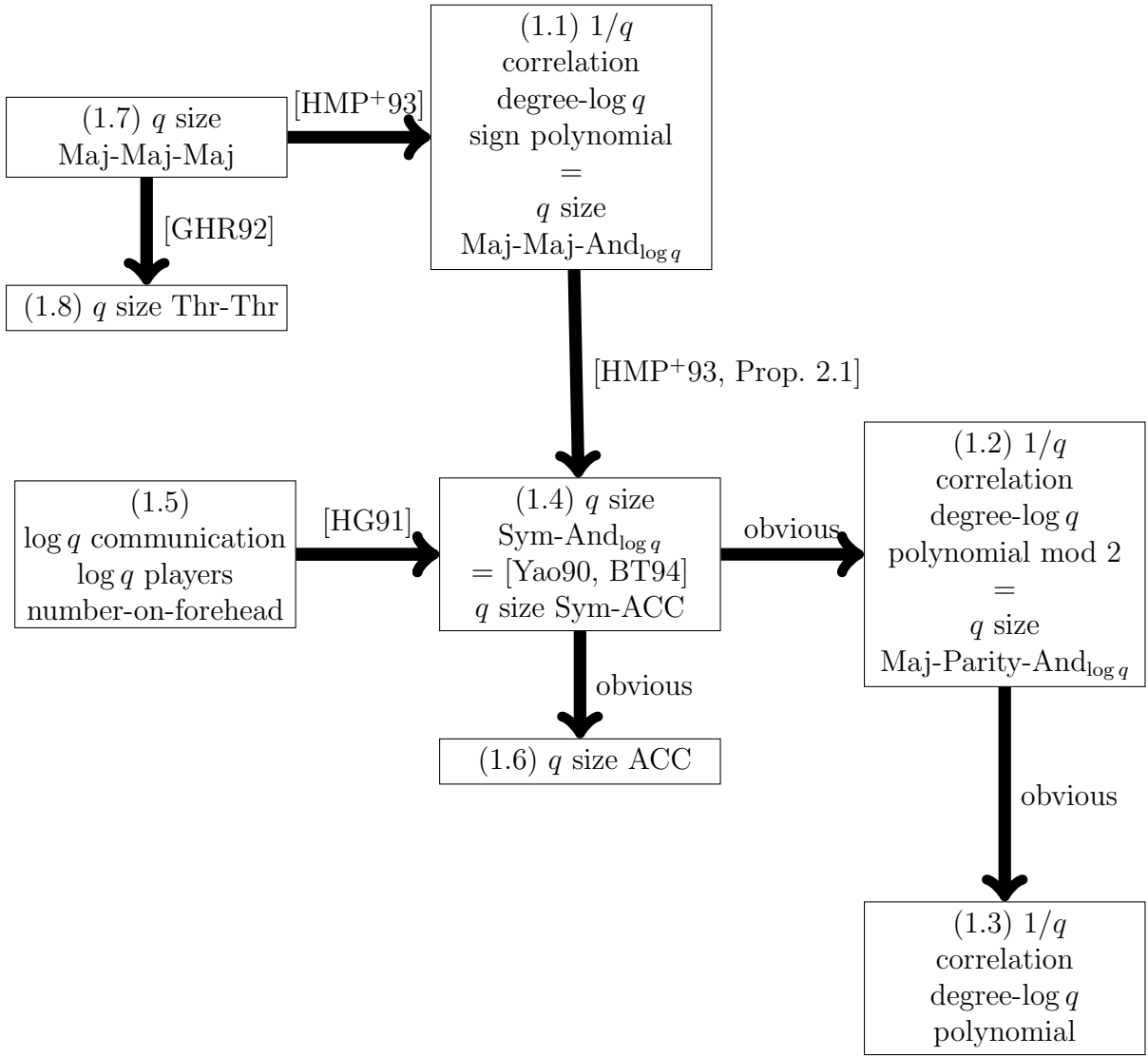
\begin{tikzpicture}[auto]

  \node [box] (corsign) {\mycon \label{cha-corsign} $1/q$ correlation\\degree-$\log q$\\ sign polynomial\\ =\\ $q$ size Maj-Maj-And$_{\log q}$};

  \node [box, below right=3cm of corsign] (cormod2) {\mycon \label{cha-cormod2} $1/q$ correlation degree-$\log q$ polynomial mod 2\\=\\ $q$ size Maj-Parity-And$_{\log q}$};

  \node [box, below=3cm of cormod2] (correal) {\mycon \label{cha-correal} $1/q$ correlation degree-$\log q$ polynomial};

  \node [box, left=2cm of cormod2] (symacc) {\mycon \label{cha-symacc} $q$ size Sym-And$_{\log q}$\\= \cite{Yao90,BeT94}\\ $q$ size Sym-ACC};

  \node [box, left=2cm of symacc] (cc) {\mycon \label{cha-cc} \\ $\log q$ communication\\ $\log q$ players number-on-forehead};

  \node [box, below=1.5cm of symacc] (acc) {\mycon \label{cha-acc} $q$ size ACC};

  \node [box, left=2cm of corsign] (majmajmaj) {\mycon \label{cha-majmajmaj} $q$ size Maj-Maj-Maj};

  \node [box, below=1.5cm of majmajmaj] (thrthr) {\mycon \label{cha-thrthr} $q$ size Thr-Thr};

  \draw [arr] (cc) -- (symacc);

  \draw [arr] (cc) -- node {\cite{HaG91}} (symacc);

  \draw [arr] (symacc) -- node {obvious} (acc);

  \draw [arr] (symacc) -- node {obvious} (cormod2);

  \draw [arr] (cormod2) -- node {obvious} (correal);

  \draw [arr] (corsign) -- node {\cite[Prop.~2.1]{HMPST93}} (symacc);

  \draw [arr] (majmajmaj) -- node {\cite{HMPST93}} (corsign);

  \draw [arr] (majmajmaj) -- node {\cite{GoldmannHR92}} (thrthr);

%  \draw [arr] (-6,4) -- node {from \eqref{cha-nc1}, if $q = n^c$} (majmajmaj);

\end{tikzpicture}
%\caption{} \label{fig-cor}
%\end{figure}

%\section*{Figure \ref{fig-cor}}

\bigskip

Each occurrence of $q$ stands for a quasipolynomial
function $2^{\log^c n}$ for a possibly different constant
$c$.  For example, Challenge \eqref{cha-cc} asks to
exhibit an explicit function $f$ such that for every
constants $c$ and $c'$ it holds that for sufficiently
large $n$ the function $f$ on inputs of length $n$ cannot
be computed by a number-on-forehead protocol among
$\log^c n$ players exchanging $\log^{c'} n$ bits.

The picture changes if $q$ stands for a polynomial
function $n^c$. In this case the three equalities in
\eqref{cha-corsign}, \eqref{cha-cormod2}, and
\eqref{cha-symacc} do not hold anymore.  Intuitively this
is because a polynomial in $n$ variables of degree $\log
n$ may have $n^{\Omega(\lg n)}$ terms.  In fact, Razborov
and Wigderson show in \cite{RaW93} $n^{\Omega(\log n)}$
lower bounds for Maj-Sym-And circuits, thus resolving one
side in each of these equalities.  Other than that, every
challenge is open even for $q = n^c$. The arrows that are
known to hold in this case are the ``obvious'' arrows
\eqref{cha-symacc}--\eqref{cha-acc} and
\eqref{cha-cormod2}--\eqref{cha-correal}, and the arrow
\eqref{cha-majmajmaj}--\eqref{cha-thrthr}, labeled
\cite{GoldmannHR92}. Finally, there are new arrows from
\eqref{cha-nc1} to \eqref{cha-majmajmaj} and to
\eqref{cha-symacc}.  For the technique yielding these new
arrows see e.g.~\cite[Lecture 8]{viola-gems09}.

Both Maj and Thr stand for gates that compute a threshold
function, i.e.~a function that given input bits
$(x_1,\ldots,x_s)$ outputs 1 iff $\sum_i c_i \cdot x_i
\ge t$, for fixed integers $c_i$ and $t$.  A circuit has
\emph{size} $s$ if it has at most $s$ gates and the
weights $c_i$ in every majority gate satisfies $|c_i| \le
s$.  We do not allow multiple edges.  Sym stands for a
gate computing a symmetric function.  And$_{\log q}$ is an And
gate of fan-in $\log q$.  Every other gate has unbounded
fan-in.  We use standard notation for composing gates.
For example Maj-Maj-And$_{\log q}$ refers to a circuit with
output gate Maj taking as input Maj gates taking as input
And gates with fan-in ${\log q}$ taking as input the input bits.

For simplicity all polynomials have integer coefficients.
By ``$\e$ correlation degree-$d$ polynomials''
\eqref{cha-correal} we refer to the set of functions $g :
\zo^n \to \zo$ such that there exists some distribution
$D$ on the inputs, and some polynomial $p$ of type $X$
such that $|\Pr_{x ~ D}[p(x) = g(x)] - \Pr_{x ~ D}[p(x)
\ne g(x)] | \ge \e$.  For \eqref{cha-cormod2} and
\eqref{cha-corsign} we take the output of the polynomial
modulo 2 or, respectively, the sign of the output.

We now elaborate further on some of the challenges:

\eqref{cha-cormod2} See the survey \cite[Chapter
    1]{viola-FTTCS09}.  The equality is obtained as
    follows.  The simulation of polynomials by
    circuits is proved via boosting \cite[Section
    2.2]{Freund95} or min-max/linear-programming
    duality \cite[Section 5]{GoldmannHR92}.  The
    other direction follows from the ``discriminator
    lemma'' of \cite{HMPST93}.

\eqref{cha-corsign} The equality is obtained by
    reasoning as for \eqref{cha-cormod2}. Since we are
    not restricting the magnitude of the polynomial's
    coefficients this would yield circuits where the
    middle gate is Thr, not Maj. However
    \cite[Theorem 26]{GoldmannHR92} shows that
    Maj-Thr = Maj-Maj up to a polynomial change in
    size.

\eqref{cha-correal} For more on this see
\cite{RazborovV-rave}.

\eqref{cha-thrthr} For a special case see
\cite{HansenP10}.

\eqref{cha-cc} For a special case for which the
    arrow continues to hold see \cite{BGKL03}.

\medskip

Arrow \eqref{cha-majmajmaj}--\eqref{cha-corsign}, labeled
\cite{HMPST93}, follows from the techniques in
\cite[Lemma 2.4]{HMPST93} which give that any Maj-Sym
circuit can be turned into a Maj-Maj circuit with a
polynomial increase.

\newpage

\section{Circuits and branching programs}

%\begin{figure}
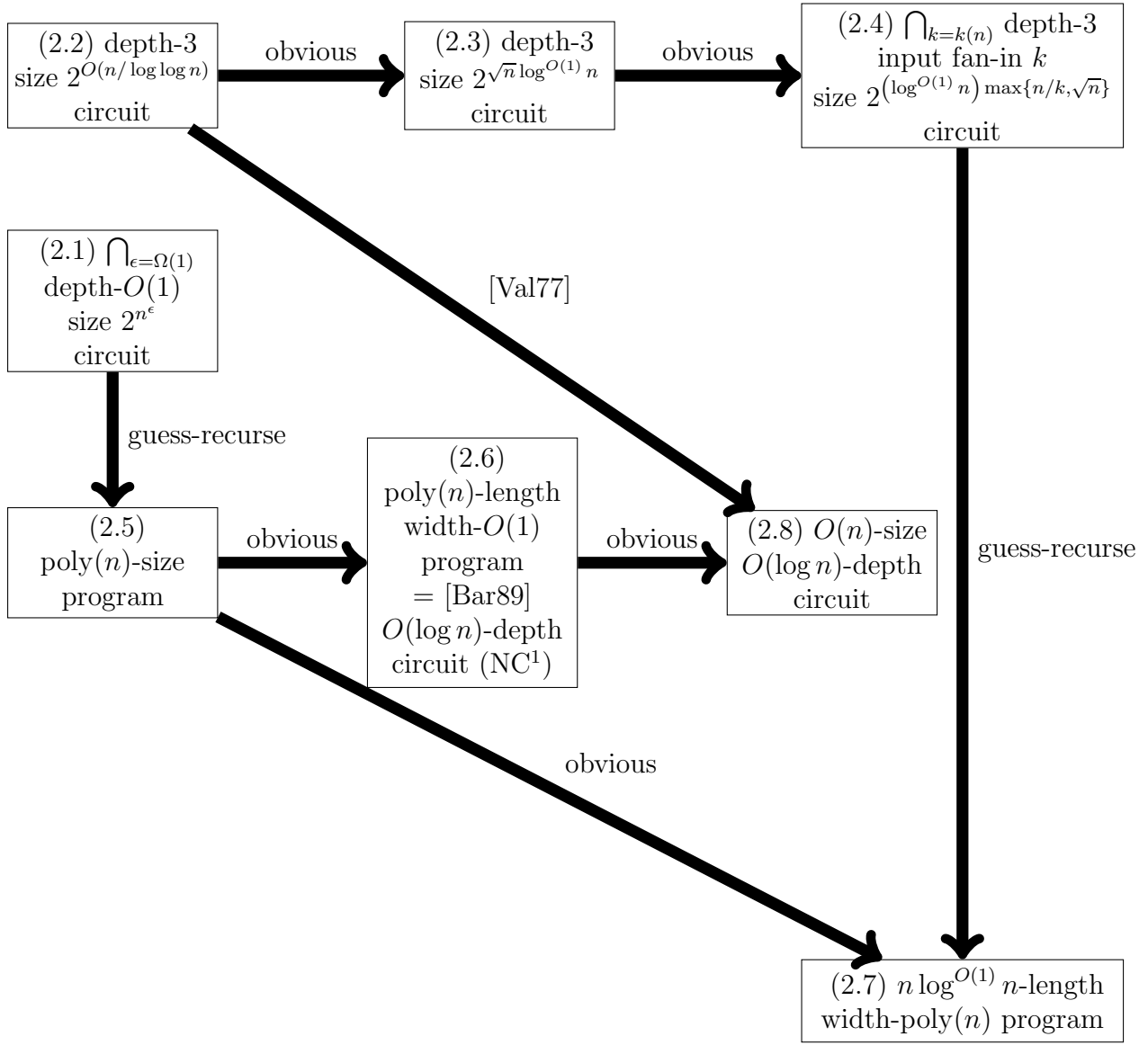
\begin{tikzpicture}[node distance = 3cm, auto]

  \node [box7] (depthd) {\mycon \label{cha-depthd} $\bigcap_{\e = \Omega(1)}$ depth-$O(1)$\\ size $2^{n^\e}$\\ circuit};

  \node [box7, above=1.5cm of depthd] (depth3valiant) {\mycon \label{cha-depth3valiant} depth-$3$\\size $2^{O(n/\log \log n)}$\\circuit};

  \node [box7, right=2.75cm of depth3valiant] (depth3sqrt) {\mycon \label{depth3sqrt} depth-$3$\\ size $2^{\sqrt{n} \log^{O(1)} n}$\\ circuit};

  \node [box11, right=2.75cm of depth3sqrt] (depth3tradeoff) {\mycon \label{cha-depth3tradeoff} $\bigcap_{k = k(n)}$ depth-$3$\\ input fan-in $k$\\ size $2^{\left( \log^{O(1)}n \right) \max\{n/k, \sqrt{n}\}}$\\ circuit};

  \node [box7, below=2cm of depthd] (brprog) {\mycon \label{cha-brprog} $\poly(n)$-size\\ program};

  \node [box7, right=2.2cm of brprog] (nc1) {\mycon \label{cha-nc1} $\poly(n)$-length\\ width-$O(1)$\\ program\\ = \cite{Barrington89}\\ $O(\log n)$-depth circuit (NC$^1$)};

  \node [box11, below=12cm of depth3tradeoff] (brprogquasilinear) {\mycon \label{cha-brprogquasilinear} $n \log^{O(1)} n$-length width-$\poly(n)$ program};

  \node [box7, right=2.2cm of nc1] (linsizelogdepth) {\mycon \label{cha-linsizelogdepth} $O(n)$-size $O(\log n)$-depth circuit};

  \draw [arr] (depthd) -- node {guess-recurse} (brprog);

  \draw [arr] (depth3tradeoff) -- node {guess-recurse} (brprogquasilinear);

  \draw [arr] (brprog) -- node {obvious} (nc1);

  \draw [arr] (brprog) -- node {obvious} (brprogquasilinear);

  \draw [arr] (nc1) -- node {obvious} (linsizelogdepth);

  \draw [arr] (depth3valiant) -- node {\cite{Val77}} (linsizelogdepth);

  \draw [arr] (depth3valiant) -- node {obvious} (depth3sqrt);

  \draw [arr] (depth3sqrt) -- node {obvious} (depth3tradeoff);

%  \draw [arr] (nc1) -- node {to \eqref{cha-majmajmaj}, if $q = n^c$} (5.25,-12);

\end{tikzpicture}
%\caption{Cap}
%\end{figure}

\bigskip

``Program'' stands for ``branching program.''
Specifically we consider layered branching programs of
width $w$ (i.e., space $\log w$) and length $t$.  The
size is $w \cdot t$.  Each node is labeled with an input
variable. The challenges remain open for the model of
oblivious branching programs where the label on each node
depends only on the layer. Recall that Nechiporuk's
argument \cite{Nechiporuk66} gives bounds of the form
$\ge n^2/\log^{O(1)} n$ on the size.  This bound gives $t
= n^2/\log^{O(1)} n$ for constant width $w = O(1)$; it
gives nothing for polynomial width $w = n^{O(1)}$.  For
polynomial or even sub-exponential width the
state-of-the-art is due to Beame, Saks, Sun, and Vee
\cite{BSSV03}.  For sub-exponential width they obtain $t
\ge \Omega(n \sqrt{\log n/\log \log n})$.

All circuits are over the basis And, Or, and Not, with
negations at the input level only.  For circuits of depth
$O(1)$ the fan-in of Or and And gates is unbounded; for
circuits of depth $\Omega(\lg n)$ the fan-in of these
gates is 2.  The \emph{size} of a circuit is its number
of edges.  Recall that for every constant $d$ the
state-of-the-art lower bounds are of the form $\ge 2^{c
n^{1/(d-1)}}$ for a constant $c$, see e.g.~\cite{Has87}.
Challenge \eqref{cha-depthd} asks to exhibit an $\e > 0$
such that for every $d$ a lower bound $2^{n^\e}$ holds.
Note for $d=3$ the state-of-the-art gives $2^{c
\sqrt{n}}$.  Challenge \eqref{depth3sqrt} asks to improve
this.  For a recent approach, see \cite{GoldreichW13}.
Further parameterized by the input fan-in $k$ of the
circuit, the available lower bounds for $d=3$ are no
better than $2^{c \max{(n/k), \sqrt{n}}}$ for a constant
$c$.  Challenge \eqref{cha-depth3tradeoff} asks to break
this tradeoff.

\medskip

The arrows \eqref{cha-depthd}--\eqref{cha-brprog} and \eqref{cha-depth3tradeoff}--\eqref{cha-brprogquasilinear}, labeled ``guess-recurse,'' are obtained via a technique
attributed to Nepomnja\v{s}\v{c}i\u{\i} \cite{Nep70}.
The arrow \eqref{cha-depthd}--\eqref{cha-brprog} continues to hold if \eqref{cha-brprog} is replaced with the functions that for every $\e > 0$ are computable by non-deterministic branching programs of length $\poly(n)$ and width $2^{n^\e}$, a class containing NL.

We give the details for the \eqref{cha-depth3tradeoff}--\eqref{cha-brprogquasilinear} arrow.
%(Cf.~e.g.~\cite[Lemma 15]{GuV04} or \cite[Lemma
%8.1]{AllenderHMPS08} for related formulations).
%Allender+ explicit in proof but not in claim
%Fan-in discussion nowhere

\begin{claim} \label{claim-branch-prog-in-depth-3}
Let $f : \zo^n \to \zo$ be computable by a branching
program with width $w$ and time $t$.  Then $f$ is
computable by a depth-3 circuit with $\le 2^{\sqrt{t \log
w}} \cdot t$ wires.  More generally, for any parameter
$b$ one can have a depth-3 circuit with
\[2^{b \log w + t/b + \log t}\]
wires, output fan-in $w^b$, and input fan-in $t/b$.
\end{claim}

The
\eqref{cha-depth3tradeoff}--\eqref{cha-brprogquasilinear}
arrow corresponds to the setting $t = n \cdot
\log^{O(1)}$ and $w = \poly(n)$.  It is obtained as
follows.  If $k \ge \sqrt{n}$ (infinitely often) the
arrow follows immediately.  If $k < \sqrt{n}$ set $b :=
t/k$ and note that the lemma gives a circuit with input
fan-in $k$ and size $\le 2^{(\log^{O(1)} n) n/k + k +
O(\log n)} \le 2^{(\log^{O(1)} n) n/k}$.

%For $w = n^{O(1)}$ and $t = n \log^{O(1)} n$, the size of
%the depth-3 circuit is $2^{\sqrt{n} \log^{O(1)} n}$.  So
%proving a depth-3 size lower bound of $2^{\sqrt{n}
%\log^{\omega(1)} n}$ implies significant progress on
%branching-program lower bounds.  Note that, for these
%parameters, depth-3 (or even depth-2) circuits compute
%more functions than branching programs.

%Even progress on lower bounds for input fan-in $k =
%\log^{O(1)} n$ would imply progress on branching-program
%lower bounds.  Indeed, for $w = n^{O(1)}$ and $t = n
%\log^{O(1)}$, for an appropriate choice of $b = n /
%\log^{O(1)} n$, the claim yields a simulation with input
%fan-in $\log^{O(1)} n$ and size $2^{n/\log^{\Omega(1)}
%n}$.

\begin{proof}
On an input $x$, guess $b$ middle points on the branching
program's computation path, at fixed times $t/b, 2t/b,
\ldots, t$.  Since the times are fixed, this is a choice
out of $w^b$.  Then verify the computation of each of the
corresponding $b$ intervals is correct.

Each interval involves paths of length $\le t/b$.  The
computation can be written as a decision tree of the same
depth.  In turn, this is a CNF with $\le 2^{t/b} t/b$
wires.

Collapsing adjacent layers of And gates we obtain a
circuit with size \[\le w^b \cdot b \cdot \le 2^{t/b} t/b
= 2^{b \log w + t/b + \log t}\] wires.

Setting $b := \sqrt{t/\log w}$ yields size
\[2^{\sqrt{t \log w} + \log t}.\]

Moreover, by construction this circuit has output fan-in
$w^b$ and input fan-in $t/b$.
\end{proof}

For an exposition of the arrow
\eqref{cha-depth3valiant}--\eqref{cha-linsizelogdepth},
labeled \cite{Val77}, see e.g.~\cite[Chapter
3]{viola-FTTCS09}.

\bibliographystyle{alpha}
\small{
\bibliography{../OmniBib}
}

\end{document}